\documentclass{article}
\usepackage{spconf,amsmath,graphicx}
\usepackage{hyperref}
\usepackage{url}
\usepackage{tabu}
\usepackage{enumitem}
\usepackage{subfigure}

\title{Subjective Quality Assessment for YouTube UGC Dataset}

\name{Joong Gon Yim, Yilin Wang, Neil Birkbeck, Balu Adsumilli}
\address{Google Inc., 1600 Amphitheatre Pkwy., Mountain View, CA, USA 94043}

\begin{document}

\maketitle

\begin{abstract}
Due to the scale of social video sharing, User Generated Content~(UGC) is getting more attention from academia and industry. To facilitate compression-related research on UGC, YouTube has released a large-scale dataset~\cite{Wang2019UGCDataset}. The initial dataset only provided videos, limiting its use in quality assessment. We used a crowd-sourcing platform to collect subjective quality scores for this dataset. We analyzed the distribution of Mean Opinion Score (MOS) in various dimensions, and investigated some fundamental questions in video quality assessment, like the correlation between full video MOS and corresponding chunk MOS, and the influence of chunk variation in quality score aggregation. 
\end{abstract}

\section{Introduction}
\label{sec:intro}

Balancing the trade-off between bitrate and visual quality is the core problem of video compression. Numerous quality metrics (SSIM \cite{Wang04SSIM} and VMAF~\cite{Li16VMAF}) have been proposed to provide accurate video quality estimation. A common assumption is that the original video is pristine, and any operation on the original (processing, compression, etc.) makes it worse. Consequently, most research measures how good the resulting video is by comparing it to the original. However, this assumption is invalidate when the original is not pristine and has artifacts. In this case, it is unclear if an encoder should spend bits to faithfully reproduce these artifacts. Given that User Generated Content (UGC) is most of the video on sharing platforms (e.g., YouTube) and most of UGC are non-pristine, efficiently handling such non-pristine originals is becoming an important and challenging research topic. However, there were few public datasets for UGC specifically, and that limits the research in this area.

Recently, YouTube released a large-scale UGC dataset~\cite{Wang2019UGCDataset} sampled from 1.5 million YouTube videos with the Creative Commons license. The dataset contains 1500 20-second video clips, covering 15 categories (e.g., Gaming, Sports, and Music Video) and various resolutions (from $360$P to $4$K). Consequently, this data is a good basis for research on the practical application of video compression and video quality assessment.

\begin{figure} [t]
\includegraphics[width=8cm, trim={0 0 0 0}, clip]{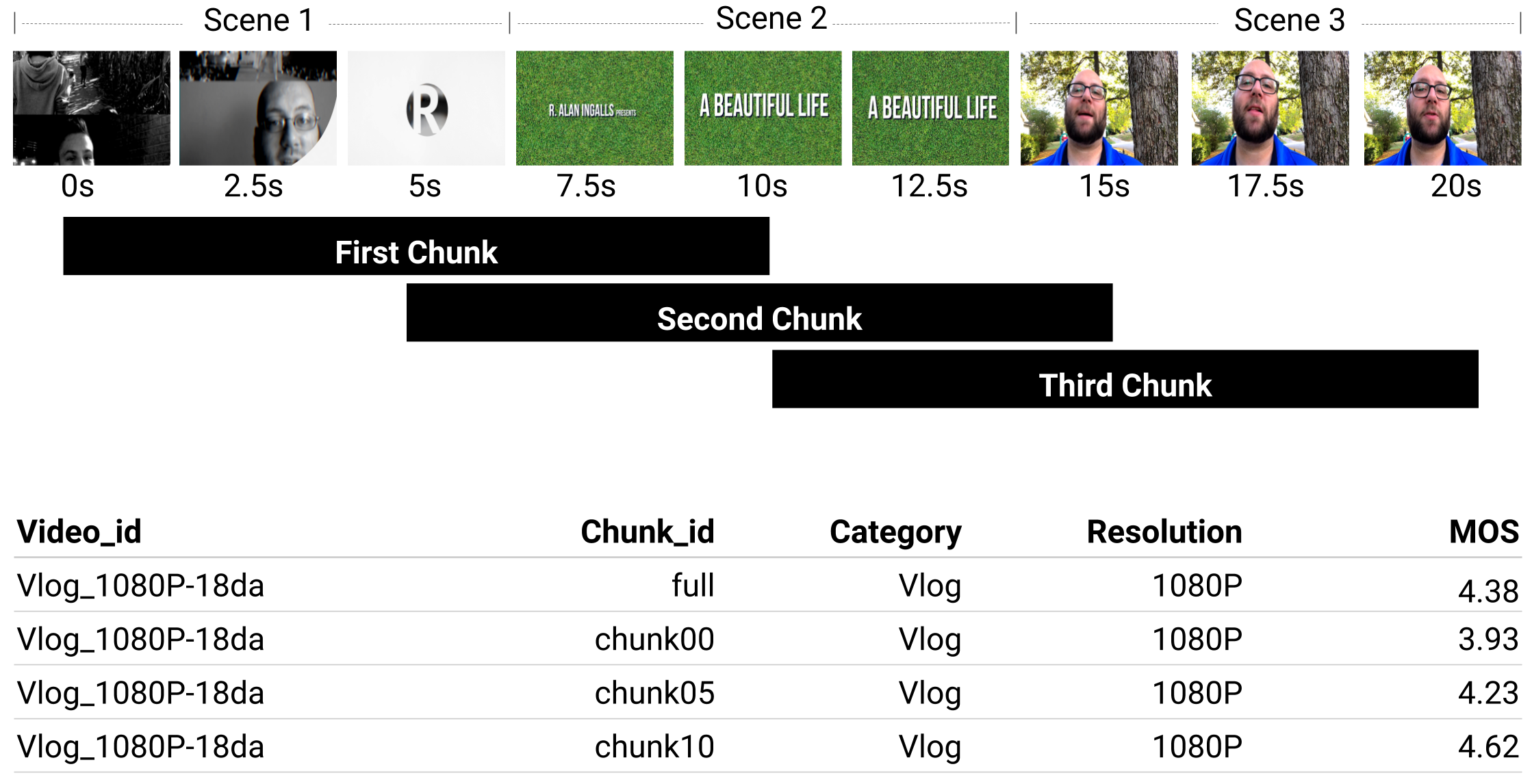}
\caption{Full and chunk MOS provided with our Dataset.}
\label{fig:website_UGC_MOS_sample}
\end{figure}

As the initial dataset provided only the raw videos, its use for quality assessment was limited. To enable novel research on compression and quality assessment on UGC, in this work we have collected and publicly released the corresponding subjective quality scores for the YouTube UGC dataset\footnote{The raw videos and corresponding Mean Opinion Scores (MOS) can be downloaded from \href{<url>}{\textit{https://media.withyoutube.com/ugc-dataset}}}. Our contributions are as follows:
\begin{itemize}[leftmargin=.15in] 
\item We use crowd-sourcing to collect hundreds of thousands of opinion scores for the complete 20-second videos in the YouTube UGC dataset, and we discuss and evaluate data cleaning strategies used to improve the quality of the released mean opinion scores (MOS) (Section~\ref{sec:Data Collection}).
\item We validate the consistency of the scores by comparison across different client/raters (e.g., age and display), and illustrate that the broad variety of original qualities in the dataset is different across content categories (Section~\ref{sec:Full Video MOS Analysis}) 
\item As UGC video often consists of multiple scenes (e.g., Fig. ~\ref{fig:website_UGC_MOS_sample}), we also collected quality scores for three overlapping 10-second chunks (starting at 0, 5, and 10 seconds). These finer grained annotations allow us to understand how the impact scene quality has on overall watching experience. 
\item With this data, we identify when aggregated chunk MOS accurately predicts full video MOS (Section ~\ref{sec:Chunk MOS Analysis}).
\item Finally, we use this dataset to indicate open challenges on UGC, where existing no-reference video quality assessment method performance is unsatisfactory  (Section~\ref{sec:Open Challenges}).
\end{itemize}

\section{Related Work} \label{sec:Related Work}

Some large-scale UGC datasets have already been released, like YouTube-8M~\cite{Haija2016YouTube8M} and AVA~\cite{Gu2018AVA}. However, they only provide extracted features instead of raw pixel data, making them less useful for compression research. 
LIVE datasets~\cite{LIVEVideoset,LIVENetflixDataset,LIVEMobileDataset} provide subjective scores collected by in-lab studies. All of them contain less than 30 individual pristine clips, along with about 150 distorted versions. Each video clip in the dataset was assessed by at least 35 people.

VideoSet~\cite{VideoSet16} contains 220 5-second clips extracted from 11 pristine videos. The target here is also quality assessment, and it provides the first three Just-Noticeable-Difference (JND) scores collected from more than 30 subjects.

The Crowdsourced Video Quality Dataset~\cite{Sinno12018CrowdsourcedQualityDataset} contains 585 10-second video clips, captured by 80 inexpert videographers. The dataset has 18 different resolutions and a wide range of quality owing to the intrinsic nature of real-world distortions. Like in our work, subjective opinions were collected from thousands of participants using Amazon Mechanical Turk.

KoNViD-1k~\cite{konvid1k} is another large-scale dataset which contains 1200 clips with corresponding subjective scores. They started from a collection of 150K videos, grouping them by multiple attributes like blur, colorfulness etc. The final set was created by a ``fair-sampling'' strategy. The subjective mean opinion scores were gathered through crowd-sourcing.

As a new large-scale dataset (1500 videos), YouTube UGC dataset is designed to reveal quality issues that may not be addressed enough in the past. It contains features complementary to existing datasets, e.g., videos are grouped by content categories, and many of them contain more than one scene. Our dataset allows people to investigate dimensions beyond pictures, and generate new insights for realistic needs of video compression and quality assessment.

\section{Data Collection}
\label{sec:Data Collection}
\subsection{Video Preparation}
\label{ssec:Data Preparation}
Videos in the UGC dataset are in RAW YUV 4:2:0 format. In order to be playable on all clients' browsers, original videos were transcoded by H.264~\cite{ieee2003h264} with Constant Rate Factor (CRF) value of 10 to preserve the quality as close as possible to the original version. As mentioned in Sec.~\ref{sec:intro}, one of our goals is to explore the relationship between full video quality and chunk quality, so besides the full 20-second version, each video was also split into three overlapping 10-second chunks, with starting offsets of 0, 5, and 10 seconds.
To avoid the impact of variance in video length, we held two separate subjective tests, one for full 20-second videos, the other for all 10-second chunks. 

\subsection{Subjective Test Platform}
\label{ssec:Subjective Test Platform}
Our subjective test platform is an interactive full-screen web application that is built on top of Amazon Mechanical Turk. 
%
%
When a subject first entered the test platform, some client information was collected and checked with the qualification requirements (e.g. not mobile devices). Any subject that failed to meet the requirements was rejected from the test. 
All valid subjects were shown three training videos in the beginning to get them familiar with the testing process. These three videos were chosen to exemplify bad, okay, and good qualities. After the training, subjects were presented testing videos that were randomly sampled from the YouTube UGC dataset and two anchor videos that all subjects watched.

For each video, subjects had to watch the entire duration of the clip and were asked the quality assessment question of \textit{How was the overall video quality?} The rating was given on a 1 to 5 scale slider, adjustable in 0.1 increments, where each integer is marked as \textit{Bad (1)}, \textit{Poor (2)}, \textit{Fair (3)}, \textit{Good (4)}, and \textit{Excellent (5)}.

After subjects completed the test, they were asked exit survey questions to provide their age, gender, occupation, vision, device type, and diagonal length of their display device.

The test was designed to be finished within 20 minutes and the average time to complete the test was 22.4 minutes. Each video clip was finally rated by more than $100$  subjects.

\subsection{Data Cleanup}
\label{ssec:Data Cleanup}
Although our selection of subjects were relatively high quality, there were inevitably some subjects that spammed the platform and should be removed before further analysis. Most existing data cleanup methods (e.g., in ITU-R Rec. BT.500) are designed for in-lab studies and assume that all subjects watched all videos, which is different from our setup. We analyze three methods for cleaning the data:
\begin{enumerate}[nosep]
    \item Remove \textbf{subjects} with low linear correlation ($<0.4$)
    \item Remove \textbf{subjects} that rated anchor videos wrong
    \item Remove \textbf{scores} outside of the middle 80 percent for the same video, which keeps median fixed.
\end{enumerate} 
As the comparison shown in Table~\ref{tab:outlierRejectionPerformance}, Raw ($80\%$) outperforms other options because it reduces the average of standard deviations for each video's scores (STD) without decreasing the average number of scores per video (Count) significantly. Chunk-level measurements share a similar story.

\begin{table}[ht]
\small
    \centering
    \begin{tabular}{| l || c | c | c |}
        \hline
        \textbf{Data}         & \textbf{STD}  & \textbf{Count} \\ \hline
        Raw (all)             & 0.83          & 145            \\ 
        Linear Correlation    & 0.81          & 127            \\
        Anchor                & 0.77          & 127            \\
        \textbf{Raw ($80\%$)} & \textbf{0.57} & \textbf{123}   \\ \hline
    \end{tabular}
    \caption{Comparison of data cleanup options.}
    \label{tab:outlierRejectionPerformance}
\end{table}

\section{Full Video MOS Analysis}
\label{sec:Full Video MOS Analysis}
\begin{figure} [ht]
\centering
\includegraphics[width=5cm, trim={5.5cm 10.5cm 6cm 10.5cm}, clip]{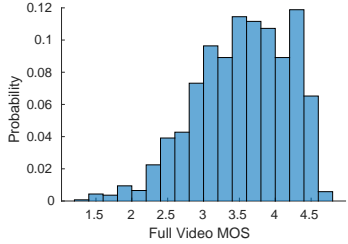}
\caption{Distribution of full video MOS.}
\label{fig:full_mos_distribution}
\end{figure}

\begin{figure} [ht]
\begin{subfigure}
\centering
\includegraphics[width=2.7cm, trim={6.5cm 11cm 7.5cm 11cm}, clip]{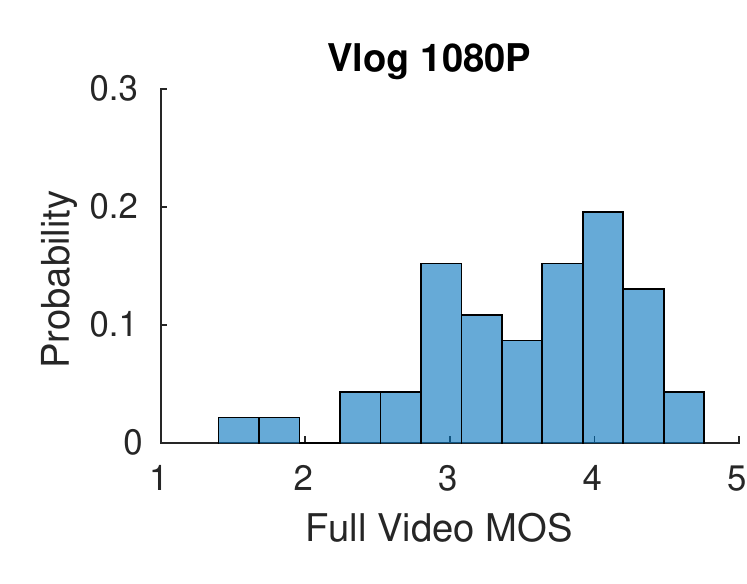}
\end{subfigure}
\begin{subfigure}
\centering
\includegraphics[width=2.7cm, trim={6.5cm 11cm 7.5cm 11cm}, clip]{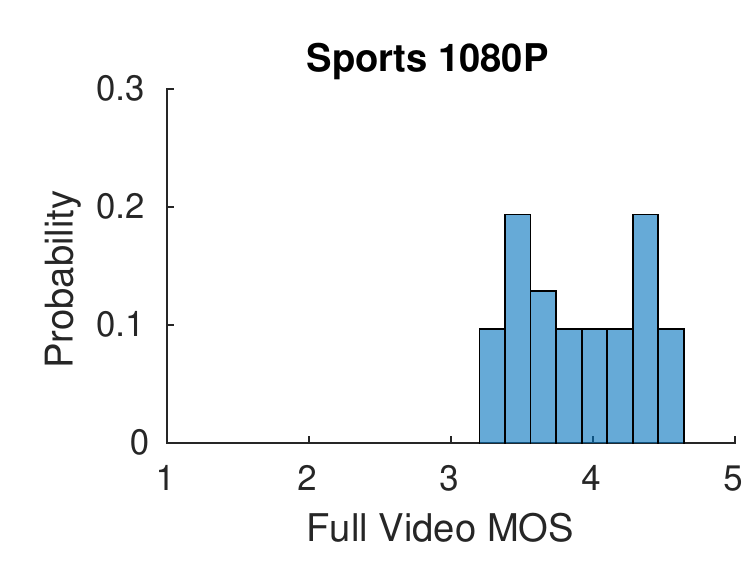}
\end{subfigure}
\begin{subfigure}
\centering
\includegraphics[width=2.7cm, trim={6.5cm 11cm 7.5cm 11cm}, clip]{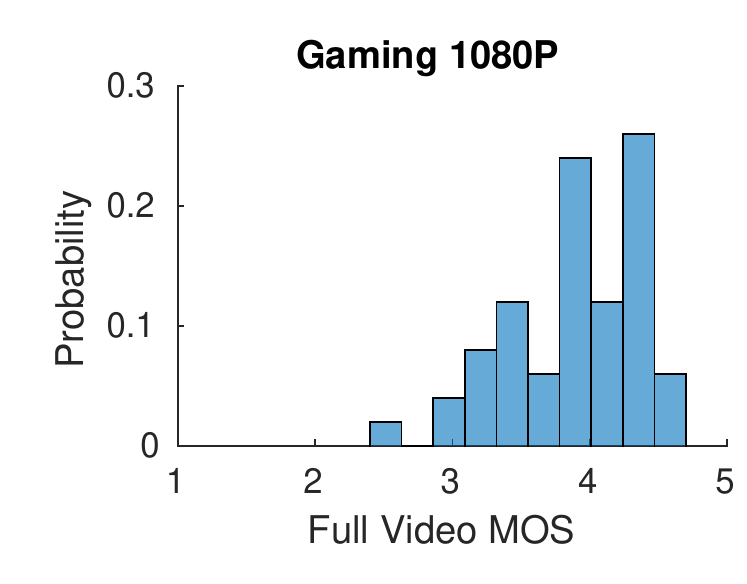}
\end{subfigure}
\caption{Distribution of full video MOS for various content categories in 1080P.}
\label{fig:category_resolution_mos}
\end{figure}

This section discusses subjective quality scores for full 20-second clips, and most conclusions also hold for chunk-level measurements. 
Fig.~\ref{fig:full_mos_distribution} shows the distribution of MOS for individual videos, where $80\%$ of videos got MOS greater or equal to $3.0$, which in some sense implies that people are satisfied with the quality for most YouTube videos. When breaking down by content category and input resolution, we find that the MOS distributions of different subsets are noticeably different. Fig.~\ref{fig:category_resolution_mos} compares three content categories: \textit{Vlog}, \textit{Sports}, and \textit{Gaming} with input resolution at 1080P, where we can see the MOS of \textit{Vlog} videos are widely spread in the entire score range, while most MOS for \textit{Sports} and \textit{Gaming} videos are greater than $3.0$ and their median MOS is around $4.0$. Such variance in content category and resolution suggests that it's promising to improve the compression efficiency by adjusting strategies based on content category and resolution.

\begin{figure} [ht]
\begin{subfigure}
\centering
\includegraphics[width=4cm, trim={0 0 0 0}, clip]{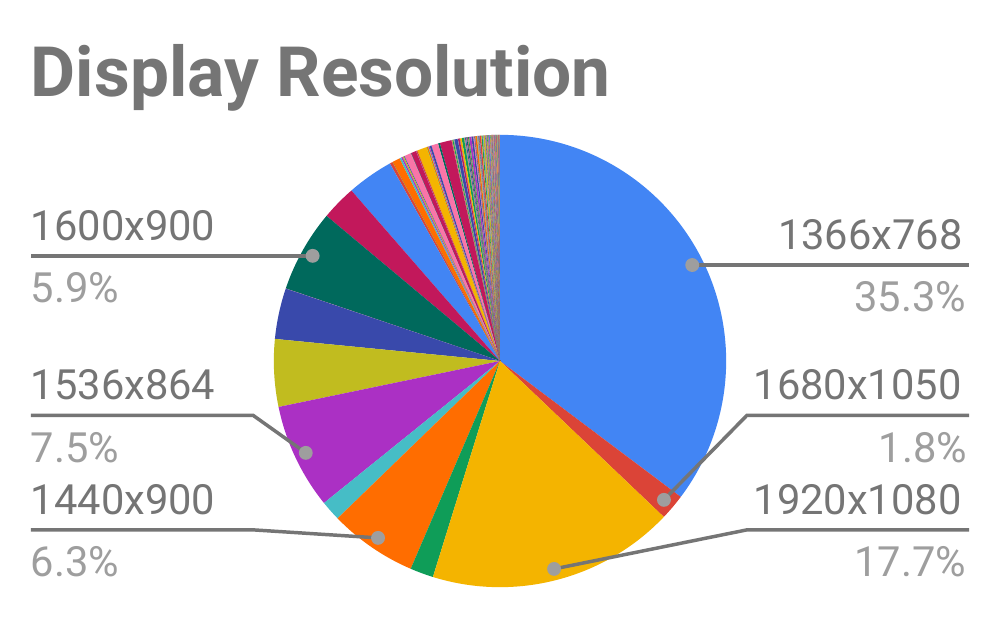}
\end{subfigure}
\begin{subfigure}
\centering
\includegraphics[width=4cm, trim={0 0 0 0}, clip]{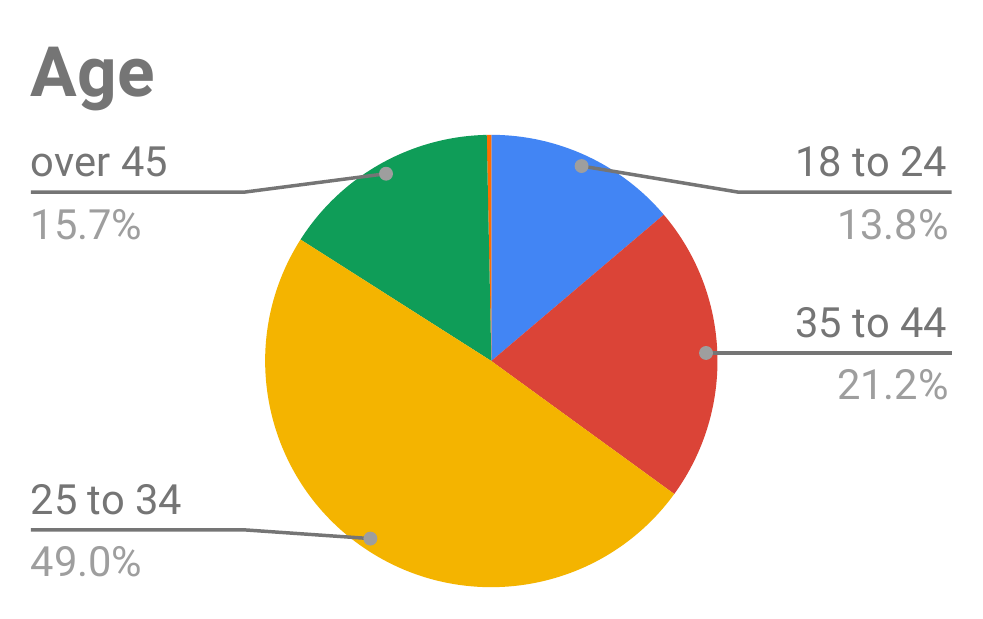}
\end{subfigure}
\caption{Distribution of age and display resolution for subjects.}
\label{fig:gender_age_disres}
\end{figure}

\begin{figure} [ht]
\begin{subfigure}
\centering
\includegraphics[width=4.0cm, trim={4.5cm 9.2cm 5cm 8.5cm}, clip]{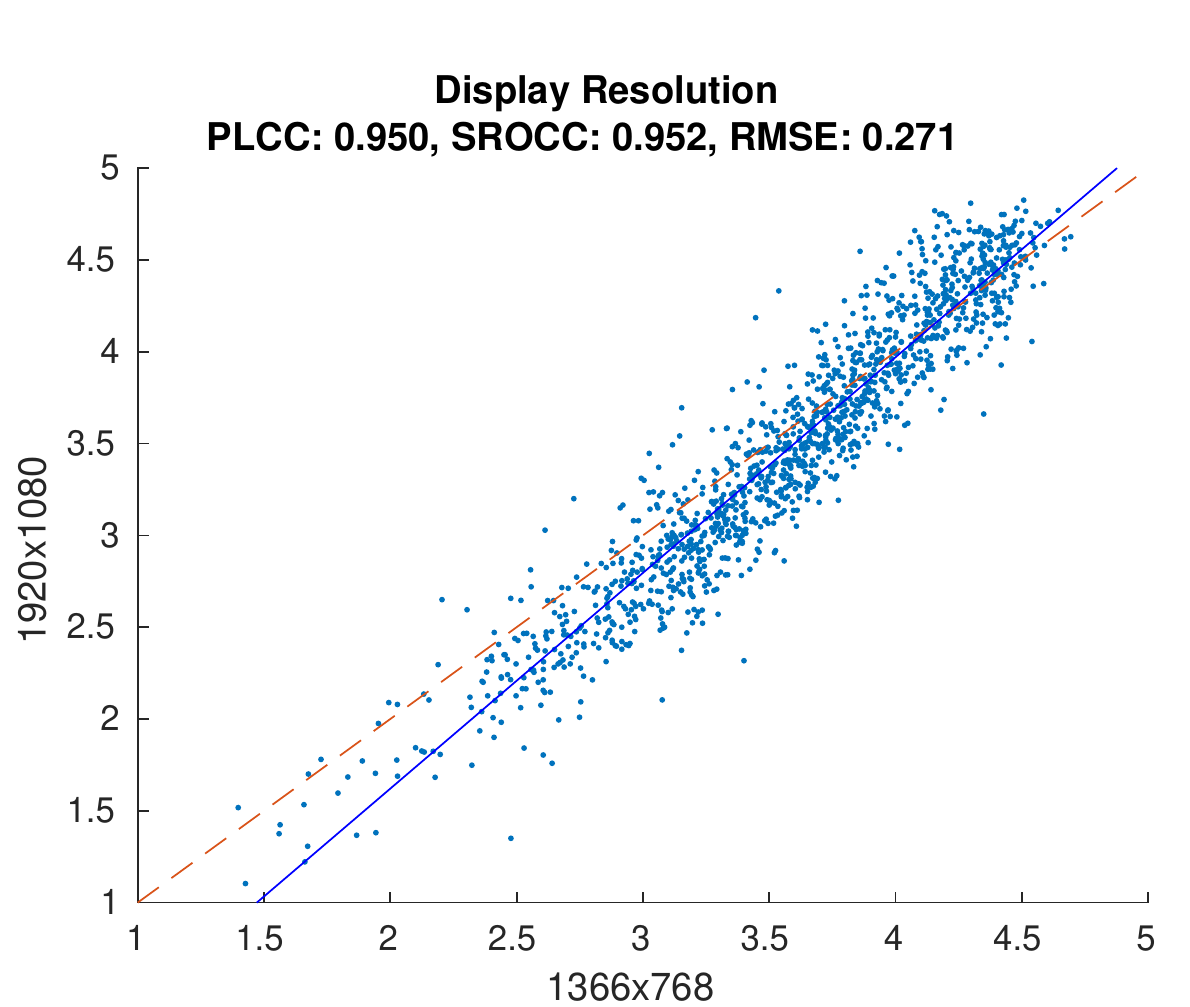}
\end{subfigure}
\begin{subfigure}
\centering
\includegraphics[width=4.2cm, trim={0 0 0 0}, clip]{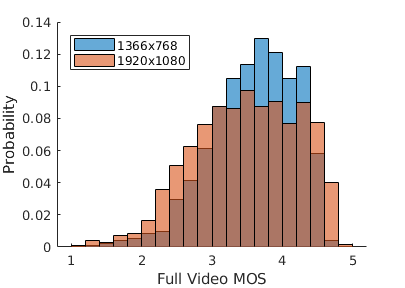}
\end{subfigure}
\caption{MOS correlation and distribution for display resolutions $1366\times{768}$ and $1920\times{1080}$. The blue line is fitted by linear regression, and the red dash line means $y=x$.}
\label{fig:mos_distribution_disres}
\end{figure}

The distributions of age and display resolution for all participants are shown in Fig.~\ref{fig:gender_age_disres}. Notice that in order to reduce the influence of devices, our subjective tests don't support cellphones, so two major display resolutions are $1366\times{768}$ ($35.3\%$) and $1920\times{1080}$ ($17.7\%$). From Fig.~\ref{fig:mos_distribution_disres} we can see these two display resolutions have high Pearson's Linear Correlation Coefficient (PLCC=$0.950$) and Spearman's Rank Correlation Coefficient (SROCC=$0.952$). The MOS spread for $1366\times{768}$ is relatively narrow w.r.t $1920\times{1080}$, and the standard deviation of all videos' MOS for $1366\times{768}$ and $1920\times{1080}$ are $0.59$ and $0.73$ respectively. Higher display resolutions show more quality variance, since low quality videos got lower scores and high quality videos got slightly higher scores on $1920\times{1080}$ display.

Table~\ref{tab:mos_plcc_age} shows the linear correlations for subjects in different age ranges, where MOS rated by subjects from different age ranges have very high correlation.
\begin{table}[ht]
\centering
\begin{tabular}{| l || c | c | c | c |}
    \hline
    Age    & \textit{25 to 34}  & \textit{35 to 44}  & \textit{over 45} \\ \hline
    \textit{18 to 24}    & 0.962     & 0.951     & 0.926     \\ 
    \textit{25 to 34}    &           & 0.974     & 0.955      \\ 
    \textit{35 to 44}    &           &           & 0.959     \\ \hline
\end{tabular}
\caption{Linear correlation (PLCC) between MOS rated by subjects within different age ranges.}
\label{tab:mos_plcc_age}
\end{table}

\section{Chunk MOS Analysis}
\label{sec:Chunk MOS Analysis}
A fundamental question in video quality assessment is how to estimate the overall video quality score from chunk (or frame) quality scores. As shown in Fig.~\ref{fig:website_UGC_MOS_sample}, a video usually consists of multiple scenes, which may change among very different context (e.g. indoor-outdoor switches, or slow and fast motions). The perceptual quality of different scenes could be significantly different. 
In this section, we will investigate the relationship between full video and chunk quality.


In practice it is common to obtain the overall video MOS by aggregating multiple chunk MOS. Three common aggregation methods are: the maximum of chunk MOS ($MOS_{max}$), the minimum of chunk MOS ($MOS_{min}$), and the average of chunk MOS $MOS_{avg}$. As shown in Table~\ref{tab:mos_full_and_chunk}, $MOS_{avg}$ has the best performance. The difference among these three aggregation methods are small, and the correlation between the full video MOS and the worst aggregation method (i.e. $MOS_{min}$) is still high (PLCC=$0.959$). In general, it is promising to use aggregated chunk MOS as the overall video MOS, and $MOS_{avg}$ is noticeable better in PLCC and RMSE than using $MOS_{max}$, $MOS_{min}$, or single chunk MOS.
\begin{table}[t]
\centering
\small
\begin{tabular}{| l || c | c | c |}
\hline
            & PLCC  & SROCC & RMSE  \\ \hline
$MOS_{max}$ & 0.968 & 0.967 & 0.172 \\
$MOS_{min}$ & 0.959 & 0.965 & 0.273 \\
$MOS_{avg}$ & 0.976 & 0.976 & 0.166 \\ \hline
\end{tabular}
\caption{PLCC, SROCC, and RMSE between full video MOS and MOS predicted from chunk MOS.}
\label{tab:mos_full_and_chunk}
\end{table}

To explore the influence of chunk MOS variation on MOS aggregation, videos are grouped by their standard deviation of chunk MOS($STD_{MOS}$).
Table~\ref{tab:mos_chunk_variance} shows the correlations between full video MOS and averaged chunk MOS for different $STD_{MOS}$. 
There are $69.1\%$ videos whose $STD_{MOS}$ is less than 0.1, $22.4\%$ in $[0,1, 0.2)$, $7.6\%$ in $[0.2, 0.5)$, and $0.9\%$ greater than $0.5$. 
Less than $10\%$ videos have large variances in chunk MOS ($STD_{MOS} \ge 0.2$), which means that most videos have consistent quality across chunks. 
We can clearly see a trend that smaller $STD_{MOS}$ has better PLCC, SROCC, and RMSE. There is a significant performance drop when $STD_{MOS}$ is greater than $0.2$, and averaged chunk MOS becomes unreliable when $STD_{MOS}$ is greater than $0.5$. 
\begin{table}[ht]
\centering
\small
\begin{tabular}{| l || c | c | c | c |}
\hline
$STD_{MOS}$  & Count & PLCC  & SROCC & RMSE  \\ \hline
$[0.0, 0.1)$ &   947 & 0.983 & 0.983 & 0.142 \\
$[0.1, 0.2)$ &   307 & 0.972 & 0.967 & 0.177 \\
$[0.2, 0.5)$ &   104 & 0.934 & 0.936 & 0.267 \\
$[0.5, inf)$ &    13 & 0.796 & 0.560 & 0.329 \\ \hline
\end{tabular}
\caption{Video count, PLCC, SROCC, and RMSE between full video MOS and averaged chunk MOS for different chunk MOS variances ($STD_{MOS}$).}
\label{tab:mos_chunk_variance}
\end{table}

\section{Open Challenges for UGC}
\label{sec:Open Challenges}
By analyzing the YouTube UGC dataset MOS, we've identified a few open-ended questions worth further investigation.

\subsection{Quality Aggregation among High Variant Chunks}
\begin{figure} [t]
\centering
\includegraphics[width=9cm, trim={0 0 0 0}, clip]{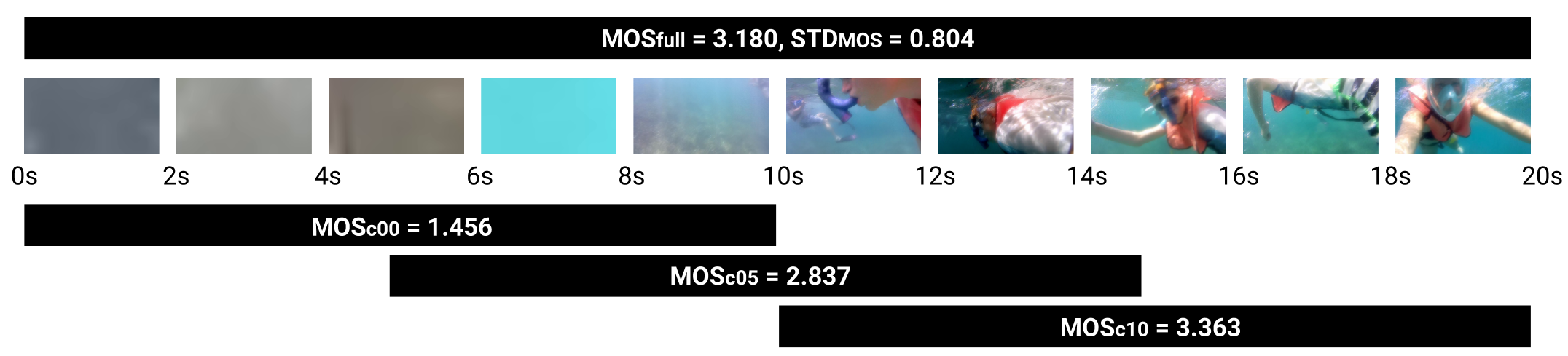}
\caption{Sample video with high chunk MOS variance, where $MOS_{full}$ ($3.18$) is much higher than $MOS_{avg}$ ($2.55$). }
\label{fig:chunk_mos_example}
\end{figure}

As shown in Table~\ref{tab:mos_chunk_variance}, when the variance of chunk quality is high, the average chunk score is an unreliable predictor of overall video quality.
Fig.~\ref{fig:chunk_mos_example} shows an example whose $STD_{MOS}=0.804$. The MOS for the first 10 seconds ($MOS_{c00}$) is low ($1.456$), mainly due to meaningless content in the first chunk. The perceptual quality significantly increased in later chunks as viewers started to see meaningful content (snorkeling boy). The overall quality ($MOS_{full}=3.180$) is very different from the averaged chunk quality ($MOS_{avg}=2.55$), suggesting that perceptual quality is mainly influenced by the meaningful part of the video. Estimating overall quality for videos with high chunk variation is still an open question, which may require high level information, like content analysis, in design.

\subsection{No-reference Quality Metric for UGC}
No-reference quality assessment has been a challenging research topic for decades, and it appears to be increasingly difficult for UGC cases. Table~\ref{tab:correlation_mos_vs_no_ref_metrics} shows the performance of some existing no-reference metrics: 
BRISQUE~\cite{Mittal2012BRISQUE}, NIQE~\cite{Mittal2013NIQE}, VIIDEO~\cite{Mittal2016VIIDEO}, SLEEQ~\cite{Ghadiyaram17SLEEQ}, and NIMA~\cite{Talebi2018NIMA}. The first four metrics are based on Natural Scene Statistics (NSS), and the last metric is trained by a Convolutional Neural Network model. All metrics are computed with their default parameters and output scores are rescaled into $[1, 5]$ using a nonlinear logistic function~\cite{Sheikh2006IAQEval}. None of these metrics has good correlations with MOS, and the possible reason is that those metrics (especially the first four) were designed for catching issues caused by compression, while some of our UGC videos have aesthetic quality issues.
\begin{table}[ht]
\centering
\small
\begin{tabular}{| l || c | c | c |}\hline
        & PLCC  & SROCC & RMSE  \\ \hline
BRISQUE & 0.112 & 0.121 & 0.639 \\
NIQE & 0.105 & 0.236 & 0.640 \\
VIIDEO & 0.146 & 0.130 & 0.637 \\
SLEEQ & 0.063 & 0.047 & 0.703 \\
NIMA  & 0.551 & 0.533 & 0.551 \\\hline
\end{tabular}
\caption{PLCC, SROCC, and RMSE between full video MOS and existing No-reference metrics.}
\label{tab:correlation_mos_vs_no_ref_metrics}
\end{table}

\section{Conclusion}
\label{sec:Conclusion}
We described how to collect the subjective data for YouTube's UGC dataset, and discussed strategies to remove noise in the raw subjective scores collected through crowd-sourcing. The full video MOS were analyzed by content category and rater context (age and display) to gain insights on the dataset. Further, the influence of chunk variation on full video MOS was also investigated. We hope the released quality scores enable new insights on UGC compression-related research.

\bibliographystyle{IEEEbib}
\bibliography{citations}

\end{document}